\documentclass[11pt]{article}
\usepackage{amsmath,amssymb,amsfonts} 
\usepackage{graphicx,times}
\usepackage{natbib}
\usepackage{mathrsfs}
\usepackage{color}
\begin{document}

\title{Vapor Transport of a Volatile Solvent 
for a Multicomponent Aerosol Droplet}

\author{James Q.~Feng}

\maketitle

\begin{center}
{Optomec, Inc., 2575 University Avenue, Suite \#135, St. Paul, Minnesota 55114, USA} \\
{{\bf email:} jfeng@optomec.com}
\end{center}

\vspace{10 mm}

\noindent\rule{160 mm}{0.5 pt}

\begin{center}
{\bf Abstract}
\end{center}

This work presents
analytical formulas derived for 
evaluating vapor transport of a volatile solvent 
for an isolated multicomponent droplet 
in a quiescent environment,
based on quasi-steady-state approximation.
Among multiple solvent components, 
only one component is considered to be
much more volatile than the rest such that other components are 
assumed to be nonvolatile 
remaining unchanged in the droplet during the process of 
(single-component) volatile solvent evaporation or condensation.
For evaporating droplet, 
the droplet size often initially decreases following the
familiar ``$d^2$ law'' at an accelerated rate.  
But toward the end, the rate of droplet size change 
diminishes due to the presence of 
nonvolatile cosolvent. 
Such an acceleration-deceleration reversal behavior 
is unique for evaporating multicomponent droplet,
while the droplet of pure solvent has an
accelerated rate of size change all the way through the end.
This reversal behavior is also reflected 
in the droplet surface temperature evolution as 
``S-shaped'' curves.
However, a closer mathematical examination of conditions 
for acceleration-deceleration reversal indicates that the 
acceleration phase may disappear when the 
amount of nonvolatile cosolvent is relatively small 
and ambient vapor pressure is relatively high.
Because the net effect of adding nonvolatile cosolvent is 
to reduce the mole fraction of the volatile solvent such that
the saturation vapor pressure is lowered, 
vapor condensation onto the multicomponent droplet is predicted to 
occur when the ambient vapor pressure is subsaturated 
with respect to that for the pure volatile solvent. 
In this case, the droplet will grow asymptotically toward a 
finite size.
But when the ambient vapor pressure becomes supersaturated 
with respect to that for the pure volatile solvent,
the condensation growth of droplet can continue indefinitely 
without bound. 

\vspace{2 mm}

\noindent\rule{160 mm}{0.5 pt}

\vspace{8 mm}


\section{Introduction}
In Aerosol Jet$^{\circledR}$ additive manufacturing,
the ink, consisting of both volatile and nonvolatile solvents besides
the solids, is atomized into a 
dense mist of fine aerosol droplets 
usually about 
$50$ nL/cc
or more.
The ink mist,  
with droplets of typical diameters in the range 
between $1$ to $5$ $\mu$m,
is then carried through conditioning devices to an 
aerodynamic focusing nozzle for 
precision material deposition onto 
a substrate \citep[cf. ][]{zollmer2006, kahn2007, christenson2011}.
The primary functionality of the volatile component 
is to reduce ink viscosity so as to enable effective atomization.
Once atomized, the ink droplets are desired to have high
solid content especially when they arrive at the substrate.
Higher solid content corresponds to higher viscosity as 
preferred for retaining the printed features at 
high mass deposition rate, which
may not be easily achieved with an ink too ``runny'' 
on the substrate.
Therefore, removal of the volatile component in
ink droplets before deposition plays an important role 
in high-quality   
Aerosol Jet$^{\circledR}$ 
printing.
In fact, the ink for 
Aerosol Jet$^{\circledR}$ 
printing is typically formulated to 
contain solvents with one volatile component
and one non-volatile component. 
The reason for having a non-volatile component
is, while the volatile component is mostly removed, 
to retain cohesivity of the printed features for 
further solidification processing 
such as controlled drying, sintering,
with appropriate annealing. 

Because the ink droplets in the 
Aerosol Jet$^{\circledR}$ 
printing process
travel through a transportation channel
from the atomizer to 
the deposition nozzle 
before arriving the substrate,
opportunities are available for 
``in-flight'' droplet solid content adjustment
via the channel shape design, 
selection of mist flow rate and sheath flow rate, 
variation of the ratio between 
volatile and nonvolatile solvent components 
in the ink formulation, etc.
Thus, non-contact deposition of ink droplets with high solid content
becomes possible with the
Aerosol Jet$^{\circledR}$ 
technology,
enabling single-pass high-aspect-ratio feature printing.
When designing-optimizing an  
Aerosol Jet$^{\circledR}$ 
printer, formulating an ink, or 
developing a process recipe for a particular application,
understanding the physical principles behind 
volatile solvent transport for 
a multicomponent ink droplet becomes practically 
important with an analytical theory 
that can provide reasonable estimate of 
relevant time scales involved therewith.

The main purpose of the present work is to 
derive analytical formulas for conveniently evaluating 
the evaporation as well as condensation characteristics of 
an isolated ink droplet.
It can be easily used to provide 
theoretical guidance to ink formulation as well as 
Aerosol Jet$^{\circledR}$ 
printing recipe development.
For consistent printing performance, 
relative motions of ink droplets in the
Aerosol Jet$^{\circledR}$ 
mist stream are minimized by design 
as the mist flowing through the deposition nozzle and 
exposing to a coflowing sheath gas with controllable solvent 
vapor contents,
where substantial droplet evaporation occurs.
Thus, a model based on vapor transport of a volatile solvent 
for an isolated multicomponent droplet in
a quiescent surrounding gas is expected to be 
quite relevant.

In a bigger picture, 
the evaporation behavior of aerosol droplets 
is of interested to a variety of applications 
such as spray drying \citep{lin2003}, 
liquid fuel combustion \citep{aggarwal1998}, 
pharmaceutical aerosol drug delivery to the lungs \citep{peng2000}, 
atmospheric science \citep{pruppacher1978}, to name a few.
It is a subject studied theoretically by 
many authors 
\citep[e.g.,][]{maxwell1878, pruppacher1978, ravindran1982, kulmala1993, 
widmann1997}.
Analytical formulas were presented for 
evaporation of a single component (unitary) droplet 
\citep{maxwell1878, pruppacher1978} 
or evaporation of a two-component (binary) droplet
with constant and uniform composition \citep{kulmala1993, vesala1997}.
In mathematical derivations, the assumptions of
``quasi-steady-state'' and ``diffusion-controlled'' 
mass transfer have typically been used.
Coupled nonlinear equations for multicomponent evaporation of 
a multicomponent droplet were derived with solutions 
computed numerically \citep{ravindran1982, widmann1997, wilms2005}.
Comparison between theory and experiment 
have been presented by \cite{wilms2005, davies2012}, among others. 

For 
Aerosol Jet$^{\circledR}$ 
printing,
the in-process
evaporation of only a single (volatile) solvent component
from a multicomponent droplet 
is of interest for most of the times.
Although the involvement of multiple solvent components in
the ink droplets makes mathematical derivations somewhat complicated,
an analytical formula can still be obtained 
and can be conveniently used for 
effective technology development.
In what follows, detailed mathematical derivation steps 
with simplifying 
assumptions are presented in section 2, 
followed by results and discussion in section 3 
for both evaporating droplet and growing droplet by condensation.
Finally, the key predictions are summarized in section 4.

\section{Mathematical Derivation}
A spherical multicomponent liquid droplet is considered here in 
a quiescent surrounding gas, with one of its solvent components
(the volatile component)
evaporating (or condensing) while the amount of other 
(nonvolatile) components remaining unchanged during the time scale 
of interest. 
The ink droplets suitable for  
Aerosol Jet$^{\circledR}$ 
printing typically have diameters in the range of 1 to 5 $\mu$m,
containing about $10\%$ volume of solids
(whose only role is considered
to passively occupy some of the droplet volume).
Thus, the size of an ink droplet 
after complete removal of the volatile
solvent should be about half of its original size, 
still much larger than the mean free path of 
the surrounding gas molecules under ambient condition.
It is therefore not unreasonable to ignore the 
Knudsen number effect.
Similar to \cite{feng2013},
the spherically symmetric
mass transfer of the volatile solvent vapor in
an incompressible continuous gas medium 
with density $\rho_g$ and diffusivity $D$ is governed by 
\citep[][p. 557]{bird1960}
\begin{equation} \label{diffusion-eqn}
\frac{\partial \rho_v}{\partial \hat{t}} 
+ v \frac{\partial \rho_v}{\partial r} 
= \frac{1}{r^2} \frac{\partial}{\partial r} 
\left(D \frac{\partial \rho_v}{\partial r}\right) 
\, \quad (\hat{R} \le r < \infty)
\, \, ,
\end{equation}
where $\rho_v$ denotes the mass concentration of 
the solvent vapor, $\hat{t}$ the time, and $r$ the radial distance
from the center of the droplet of radius $\hat{R}(\hat{t})$.
To satisfy the equation of continuity while accounting for
the effect of volume change during the solvent phase change,
the radial velocity $v$ of spherically symmetric incompressible
flow is simply
\citep{scriven1959} 
\begin{equation} \label{continuity-eqn}
v = \left(1 - \frac{\rho_l}{\rho_g}\right) 
\left(\frac{\hat{R}}{r}\right)^2 \frac{d \hat{R}}{d \hat{t}} 
\, \, ,
\end{equation}
where $\rho_l$ denotes the liquid density of the solvent 
under consideration.

At the droplet surface, the mass balance based on 
Fick's first law of binary diffusion accounting for the 
bulk flow effect with the gas medium flux being ignored 
\citep[][p. 502]{bird1960} leads to
\begin{equation} \label{mass-balance}
\frac{d \hat{R}}{d \hat{t}} = \frac{\rho_g D}{\rho_l (\rho_g - \rho_{v, R})}
\left(\frac{\partial \rho_v}{\partial r}\right)_{r = \hat{R}} 
\, \, ,
\end{equation}
where $\rho_{v, R}$ denotes the vapor density at droplet surface
($r = \hat{R}$).

For most volatile solvents used in 
Aerosol Jet$^{\circledR}$ 
inks,
the density of solvent vapor is much less than
that of gas medium, i.e., $\rho_{v, R} << \rho_g$ 
and the diffusivity $D$ may be treated as a constant.
Thus, in (\ref{diffusion-eqn}) the convection term
\[
v \frac{\partial \rho_v}{\partial r} 
= \left(1 - \frac{\rho_l}{\rho_g}\right) 
\left(\frac{\hat{R}}{r}\right)^2
\frac{\rho_g D}{\rho_l (\rho_g - \rho_{v, R})}
\left(\frac{\partial \rho_v}{\partial r}\right)_{r = \hat{R}} 
\frac{\partial \rho_v}{\partial r}
\sim - \frac{\rho_{v, R}}{\rho_g} D  
\frac{\hat{R}}{r^2}
\frac{\partial \rho_v}{\partial r}
\] 
may be assumed negligible comparing with other terms.
Then, with the terms associated to $\rho_{v, R}/\rho_g$ ignored, 
the solution to (\ref{diffusion-eqn}) 
leads to mass balance at the droplet surface  
\begin{equation} \label{dRdt0}
\frac{d \hat{R}}{d \hat{t}} = - \frac{D (\rho_{v, R} - \rho_{v, \infty})}{\rho_l}
\frac{1}{\hat{R}} \left(1 + \frac{\hat{R}}{\sqrt{\pi D t}}\right)
\, \, ,
\end{equation}
where $\rho_{v, \infty}$ denotes 
the ambient vapor density at $r \to \infty$.
Although exact solution to (\ref{dRdt0}) may be derived
\citep[cf.,][]{feng2013},
retaining the complexity due to the 
transient diffusion flux term $\hat{R}/\sqrt{\pi D \hat{t}}$ 
seems to be unnecessary 
because significant change of $\hat{R}$ occurs on a time scale 
$\sim \rho_l R_0^2/(\pi D \rho_{v, R})$ 
(with $R_0$ denoting the droplet initial radius 
used here as the characteristic length)
which is much greater than
$\hat{R}^2 / (\pi D)$ for $\rho_l >> \rho_{v, R}$ 
(i.e., liquid density is much greater than gas density).
Hence, the quasi-stationary mass balance equation (\ref{dRdt0}) 
may be reasonably approximated by a quasi-steady-state equation
\begin{equation} \label{dRdt1}
\frac{d \hat{R}^2}{d \hat{t}} = - \frac{2 D (\rho_{v, R} - \rho_{v, \infty})}{\rho_l}
\, \, ,
\end{equation}
which has the same form as standard expression based on 
Maxwell continuum transport approximation 
\citep[e.g.,][]{pruppacher1978, widmann1997}. 

As expected 
\citep[][p. 417]{pruppacher1978}, 
vapor transport to and from a droplet also involves substantial 
heat transfer owing to the release or absorption of 
heat of phase change, namely the latent heat. 
The resulting temperature difference between the droplet and its 
surrounding gas medium causes a flow of sensible heat 
by thermal diffusion. 
With the quasi-steady-state approximation, energy balance at 
the droplet surface leads to  
\begin{equation} \label{qssEnergy}
(\rho_{v, R} - \rho_{v, \infty})
= \frac{k}{L D} (T_{\infty} - T_R)
\, \, ,
\end{equation}
where $k$ is the thermal conductivity of the surrounding gas medium,
$L$ the latent heat due to volatile solvent phase change,
$T_{\infty}$ and $T_R$ the temperature values at $r \to \infty$ and 
$r = \hat{R}$.
Moreover, it is usually reasonable to take $\rho_{v, R}$ as
the local saturated vapor density at the droplet surface 
$\rho_{sv, R}$ (with ``$sv$'' in the subscript denoting 
saturated vapor).
Making use of the Clausius-Clapeyron relation, 
for characterizing a discontinuous phase transition between liquid
and gas, yields an approximate equation
\citep[as justified in ][ p. 420]{pruppacher1978}
relating local saturated vapor density difference between 
$r = \hat{R}$ and $r \to \infty$ to $(T_R - T_{\infty})$ as
\begin{equation} \label{rhosv}
(\rho_{sv, R} - \rho_{sv, \infty})
= \frac{\rho_{sv, \infty}}{T_{\infty}} 
\left(\frac{L}{\mathscr{R}_v T_{\infty}} - 1\right) (T_R - T_{\infty})
\, \, ,
\end{equation}
where $\mathscr{R}_v$ denotes the (specific) gas constant 
for the volatile solvent vapor (which can be calculated as 
the universal gas constant, 
$8.3143 \times 10^3$ J deg$^{-1}$ kmol$^{-1}$, divided by
the solvent molecular weight in kg kmol$^{-1}$.)

Combination of (\ref{qssEnergy}) and (\ref{rhosv}) leads to
\begin{equation} \label{rhodiff}
(\rho_{sv, R} - \rho_{v, \infty})
= \left(1 - \frac{\rho_{v, \infty}}{\rho_{sv, \infty}}\right)
\left[\frac{1}{\rho_{sv, \infty}} + \frac{L D}{k T_{\infty}} 
\left(\frac{L}{\mathscr{R}_v T_{\infty}} - 1\right)\right]^{-1}
\, \, .
\end{equation}
Thus, expressing $(\rho_{v, R} - \rho_{v, \infty})$ in terms
of corresponding vapor pressure 
$p_{v, \infty} = \rho_{v, \infty}/(\mathscr{R}_v T_{\infty})$, 
$p_{sv, \infty} = \rho_{sv, \infty}/(\mathscr{R}_v T_{\infty})$, and $T_{\infty}$ 
yields a more convenient form of (\ref{dRdt1}) as 
\begin{equation} \label{dRdt2}
\frac{d R^2}{dt} = - \frac{2}{\pi}
\left(1 - \frac{p_{v, \infty}}{p_{sv, \infty}}\right)
\left[\frac{\rho_l \mathscr{R}_v T_{\infty}}{p_{sv, \infty}} 
+ \frac{\rho_l L D}{k T_{\infty}} 
\left(\frac{L}{\mathscr{R}_v T_{\infty}} - 1\right)\right]^{-1}
\, \, ,
\end{equation}
where all the variables become dimensionless with
$R(t)$ measured in
units of the droplet 
initial (dimensional) radius $R_0$ and $t$ in units
of $R_0^2/(\pi D)$, 
i.e., $R \equiv \hat{R}/R_0$ and $t \equiv \hat{t} \pi D/R_0^2$.
Clearly, $R(t)$ decreases with $t$ when $p_{v, \infty}/p_{sv, \infty} < 1$ 
as the solvent is evaporating from the droplet.
If the droplet is in a supersaturated environment
when $p_{v, \infty}/p_{sv, \infty} > 1$,
$R(t)$ would increase with $t$ 
because in this case the solvent is condensing onto the droplet. 

If the droplet of multicomponent solution contains 
only one volatile component,
as
Aerosol Jet$^{\circledR}$
inks are typically formulated,
the saturated vapor pressure $p_{sv, \infty}$ for the volatile solvent
becomes a function of its mole fraction $x$ in the solution.
Assuming the multicomponent mixture is 
an ideal solution such that Raoult's law is applicable,
we have
\begin{equation} \label{psv}
p_{sv, \infty} = x p*_{sv, \infty}
\, \, ,
\end{equation}
with $p*_{sv, \infty}$ denoting the saturated vapor pressure of 
the pure volatile solvent.
Substituting (\ref{psv}) to (\ref{dRdt2}) 
incurs complexity due to the fact that $x$ is a function of $R(t)$.
With the (dimensionless) volume 
of nonvolatile component of liquid solution in the 
ink droplet denoted by
$4 \pi R_n^3 / 3$ and the total volume 
remaining in the droplet
when the volatile 
component is completely removed---the remnant volume---by 
$4 \pi R_r^3 / 3$ (which includes both solids and 
nonvolatile component of liquid solution),
the volume of volatile solvent component should be
$4 \pi (R^3 - R_r^3) / 3$ 
and 
\begin{equation} \label{x-eqn}
\frac{1}{x(t)} = 1 +   
\frac{\rho_n M_v}{\rho_v M_n}
\frac{R_n^3}{R(t)^3 - R_r^3}
\, \, ,
\end{equation}
where $\rho_n$ and $\rho_v$,
$M_n$ and $M_v$ denote the liquid solvent
densities, molecular weights of the nonvolatile and
volatile components, respectively.
Clearly, the mole fraction 
$x$ is generally a function of time $t$ 
during the volatile solvent evaporation,
because $R(t)$ changes with time 
while $R_n$ and $R_r$ remaining as constants.

Incorporating (\ref{psv}) and (\ref{x-eqn}) in (\ref{dRdt2})
yields
\begin{equation} \label{dRdt3}
\frac{dR^2}{dt} = - \frac{2}{\pi} \frac{
1 - {p_{v, \infty}}/{p*_{sv, \infty}}
}
{A + B / 
\left(R^3 - R_e^3\right)
}
\, \, ,
\end{equation}
where
\begin{equation} \label{Rr3}
R_e^3 \equiv R_r^3 
+ \frac{\rho_n M_v R_n^3 p_{v, \infty}}
{\rho_v M_n (p*_{sv, \infty} - p_{v, \infty})}
\, \, ,
\end{equation}
\begin{equation} \label{Adef}
A \equiv 
\frac{\rho_l L D}{k T_{\infty}}
\left(\frac{L}{\mathscr{R}_v T_{\infty}} - 1\right) + 
\frac{\rho_l \mathscr{R}_v T_{\infty}}{p*_{sv, \infty}}
\, \, ,
\end{equation}
and
\begin{equation} \label{Bdef}
B \equiv 
\frac{\rho_n M_v R_n^3}{\rho_v M_n}
\left(\frac{\rho_l \mathscr{R}_v T_{\infty}}{p*_{sv, \infty}}
+ \frac{A \, p_{v, \infty}}{p*_{sv, \infty} - p_{v, \infty}}\right)
\, \, .
\end{equation}
An analytical solution to (\ref{dRdt3}),
though apparently complicated, can be written as 
\begin{multline} \label{dropRvst}
R^2 
= 1 - \frac{2 t (1 - p_{v, \infty}/p*_{sv, \infty})}{\pi A} + 
\\
+ \frac{B}{3 A R_e} \left\{
\ln\left[ \frac{(1 - R_e)^3(R^3 - R_e^3)}{(R - R_e)^3 (1 - R_e^3)}\right]
- 2 \sqrt{3} \left(\tan^{-1} \frac{2 R + R_e}{\sqrt{3} \, R_e}
- \tan^{-1} \frac{2+ R_e}{\sqrt{3} \, R_e}\right) \right\}
\, \, .
\end{multline}
Unfortunately,
(\ref{dropRvst}) is in an implicit form of $R$ as a function $t$.
However, $t$ can be expressed as an explicit function of $R$,
which make it 
straightforward to calculate the value of $t$
from a given value of $R$ for generating curves of $R$ versus $t$.

\section{Results and Discussion}
Without going to the solution, an examination of
(\ref{dRdt3}) suggests the possible situation for 
$d R^2/d t \to 0$ as $R \to R_e$
with $R_e$ denoting the equilibrium droplet radius
as defined in (\ref{Rr3}) for 
the droplet at thermodynamic equilibrium with 
ambient vapor pressure $p_{v, \infty}$.
Of course, this can only happen 
in the presence of the nonvolatile component, i.e., 
when $R_n$ is nonzero such that $B$ is nonzero.
In the absence of the nonvolatile cosolvents (i.e., $R_n = 0$)
that can influence the equilibrium vapor pressure
of the volatile solvent, 
$B$ disappears and 
$R^2$ simply becomes a linear function of $t$ for a droplet 
containing a single 
solvent. 

While 
(nondimensional)
 $R_n$ and $R_r$
are subjected to constraint 
as $R_n \le R_r < 1$ (measured in units of 
droplet initial radius 
$R_0$),
there is no upper limit to the equilibrium radius $R_e$. 
According to (\ref{Rr3}), however,
$R_e > 1$ can only happen when 
$R_n^3 p_{v, \infty} \ne 0$
in a subsaturated environment,
i.e., $p_{v, \infty} \le p*_{sv, \infty}$, 
while satisfying the condition
$p*_{sv, \infty}/p_{v, \infty}$ 
$< 1/x(0)$ where 
\begin{equation} \label{x0}
\frac{1}{x(0)}= 1 + 
\frac{\rho_n M_v R_n^3}{\rho_v M_n (1 - R_r^3)}
\, \, .
\end{equation}
In the case of $p_{v, \infty} = x(0) p*_{sv, \infty}$, 
we have $R_e = 1$ and hence 
$d R^2/dt = 0$ forever because 
the ambient vapor pressure equals the saturation value
starting from $t = 0$.
Volatile solvent evaporation occurs when $R_e < 1$,
which makes $R(t)$ ($> R_e$) 
decreases with time until $R(t) \to R_e$.  
Vapor condensation onto the droplet
can occur either when $R_e > 1$ 
such that $R(t)$ ($< R_e$) increases with time 
until reaching $R_e$,
or when $p_{v, \infty} > p*_{sv, \infty}$.
In a supersaturated environment, 
namely, $p_{v, \infty} > p*_{sv, \infty}$,
we can have
$R_e < 0 \le R_r < 1$ according to (\ref{Rr3}),
in which case the negative $R_e$ loses its physical meaning, 
but
$d R^2/dt > 0$ as the droplet is 
growing due to $p_{v, \infty}/p*_{sv, \infty} > 1$ 
as indicated by (\ref{dRdt3}).
Supersaturated vapor will condense onto any surfaces, 
such as the channel walls,
besides the surface of a multicomponent droplet.
Noteworthy here is that (\ref{Rr3}) 
also suggests an indefinte growth of droplet (without bound) 
as $p_{v, \infty} \to p*_{sv, \infty}$,
because it leads to 
$R_e \to \infty$.  

For illustrations in this section,
water and butanol are used as realistic examples
for the volatile solvent,
and ethylene glycol (EG) for the nonvolatile solvent.
They have actually been used in several 
Aerosol Jet$^{\circledR}$
ink formulations.
Table 1 shows the typical parameters for those solvents, 
where only the liquid solvent density $\rho_l$ 
and molecular weight $M_w$ 
are relevant for the nonvolatile ethylene glycol.
The ambient temperature $T_{\infty}$ is assumed to be $300$ K
and ambient pressure to be $10^5$ Pa.
Thus, the values of $L/(\mathscr{R}_v T_{\infty})$ for water and butanol 
are $16.3$ and $21.0$, respectively.
As a consequence, the value of $\rho_{sv, R}$ is generally 
less than $\rho_{sv, \infty}$ according to (\ref{rhodiff}).
The values of $\rho_{v, R}/\rho_g$ in (\ref{mass-balance}), 
which is less than $p_{sv, \infty}/10^5$, 
are therefore $< 0.03$ and $0.0055$ for water and butanol,
as {\em a posteriori} verification 
of negligible terms associated with factor 
$\rho_{v, R}/\rho_g$ in our mathematical derivation.

\begin{table}
\caption{Solvent parameter values 
under standard ambient condition for density $\rho_l$ 
(kg m$^{-3}$), molecular weight $M_w$ (kg kmol$^{-1}$),
saturated vapor pressure $p*_{sv, \infty}$ (Pa), 
latent heat of vaporization $L$ (J kg$^{-1}$),
diffusivity in air $D$ (m$^2$ s$^{-1}$),
and dimensionless $A$ defined in (\ref{Adef}) 
using the thermal conductivity in air 
$k = 0.024$ (J m$^{-1}$ s$^{-1}$ K$^{-1}$).}
\begin{center}
\begin{tabular*}{0.75\textwidth}{@{\extracolsep{\fill}} lcccccc}
\hline
\hline
\\ 
  Solvent & $\rho_l$ & $M_w$ & $p*_{sv, \infty}$ & $L$ & $D$ & $A$ \\
\hline
\\
  water & $1000$ & $18.02$ & $2.99 \times 10^3$ & $2.26 \times 10^6$ & $2.50 \times 10^{-5}$ & $1.67 \times 10^5$ \\
  butanol & $810$  & $74.12$ & $5.53 \times 10^2$ & $7.08 \times 10^5$& $8.00 \times 10^{-6}$ & $6.21 \times 10^4$ \\
  EG & $1113$ & $62.07$ & $8.00$ & --- & --- & --- \\
\hline
\hline
\end{tabular*}
\end{center}
\end{table}

From (\ref{dropRvst}) it is clear that the effects of 
nonvolatile cosolvent are proportional to the value of $B/A$.
While the value of $A$ remains constant for a given solvent,
the value of $B$, according to (\ref{Bdef}), is 
proportional to the volume fraction of nonvolatile cosolvent
$R_n^3$ (among other factors such as the 
volatile solvent molecular weight $M_v$, etc.) 
Thus, a simple inspection of the physical property values in 
Table 1 would suggest that   
a butanol-EG droplet should exhibit
much more significant effects of nonvolatile solvents 
than a water-EG droplet.

\subsection{Evaporating Droplet}
When the equilibrium radius $R_e < 1$, 
the volatile solvent will evaporate from the droplet
of initial $R = 1$ 
until $R \to R_e$.
In the absence of the nonvolatile cosolvent, i.e., $R_n = 0$, 
we have $B = 0$ and (\ref{dropRvst}) recovers the familiar 
classical ``d$^2$ law'' for evaporation of a single component droplet
with a constant $d R^2/dt$.
In this case,
the volatile solvent can be completely removed in a finite time
\begin{equation} \label{t0}
t_0 = \frac{\pi A (1 - R_r^2)}{2 (1 - p_{v, \infty}/p*_{sv, \infty})}  
\, \, .
\end{equation}

A nonzero $R_n$ leads to a nonzero $B$ 
that causes a fundamental change in
the evaporation behavior:
the volatile solvent cannot be completely removed in a finite time
because now $d R^2/dt \to 0$ 
as $R \to R_e$ due to diminishing mole fraction of the 
volatile solvent $x$ in the droplet. 
As a consequence, it is expected to take 
forever to remove the last bit of 
the volatile solvent from a droplet in the presence of 
a nonvolatile cosolvent.

\begin{figure}[t!] \label{dropEvap1}
\includegraphics[scale=0.7]{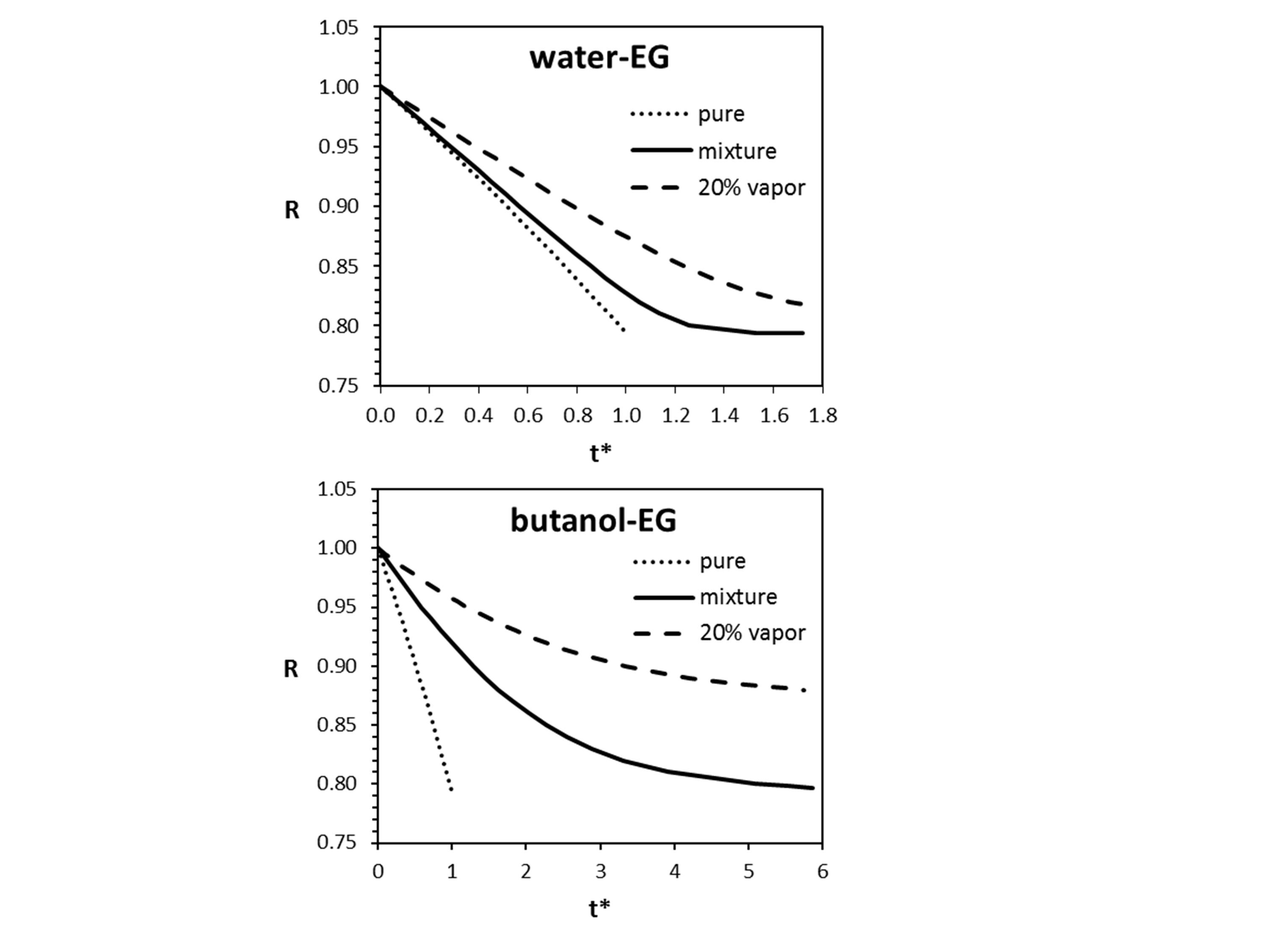}
\caption{Evaporating droplet radius evolution with time
for water-EG and butanol-EG mixtures,
with ethylene glycol (EG) serving as the nonvolatile
cosolvent with a volume fraction of
$R_n^3 = 0.4$.
The remnant volume of droplet (excluding that of 
the volatile solvent) is denoted as $4 \pi R_r^3/3$
here in this figure with the remnant volume fraction $R_r^3 = 0.5$.
The solid curve and dashed curve correspond to 
$p_{v, \infty}/p*_{sv, \infty} = 0$ and $0.2$,
with $t^*_{101} = 1.23$ and $1.71$ for water-EG droplet
while $t^*_{101} = 4.81$ and $5.49$ for butanol-EG droplet.
The dotted curve is for pure volatile solvent with 
$R_n^3 = 0$ and $R_r^3 = 0.5$ at
$p_{v, \infty}/p*_{sv, \infty} = 0$, as a reference.}
\end{figure}

If the droplet has 
a volume fraction of nonvolatile cosolvent $R_n^3 = 0.4$ and 
remnant volume fraction $R_r^3 = 0.5$
(namely, $R_r = 0.794$ as the remnant radius), 
curves of $R(t)$ are shown in Fig. 1 for evaporating 
droplets of water-EG mixture and butanol-EG mixture
with $10\%$ solids and $40\%$ nonvolatile solvents in the initial volume, 
at $p_{v, \infty}/p*_{sv, \infty} = 0$ (solid curve) 
and $0.2$ (dashed curve) which is about $25\%$
of $x(0)$ ($\approx 0.795$) for the water-EG droplet 
and nearly $50\%$ of $x(0)$ ($\approx 0.432$) 
for the butanol-EG droplet.
As a reference, 
also shown in Fig. 1 is 
the dotted curve representing the case for $R_n = 0$
and $R_r^3 = 0.5$ 
at $p_{v, \infty}/p*_{sv, \infty} = 0$ 
(according to the ``d$^2$ law'') for a single solvent droplet.
For convenience of displaying results,
$t^*$ in all figures here is used to denote
a normalized time with the value of $t_0$ in (\ref{t0})
at $p_{v, \infty}/p*_{sv, \infty} = 0$ such as  
\begin{equation} \label{time}
t^* \equiv \frac{2 t}{\pi A (1 - R_r^2)} 
\, \, .
\end{equation}
Accordingly,
the dimensional time corresponding to $t^* = 1$ can be calculated as 
\begin{equation} \label{dimensional-time}
\frac{A (1 - R_r^2) R_0^2}{2 D} \,\, , 
\end{equation}
with $R_0$ and $D$ taking appropriate dimensional values.  
For example, a water droplet of $R_0 = 10^{-6}$ m 
(one micron radius, as the typical mean ink droplet radius in the
Aerosol Jet$^{\circledR}$
mist)
containing $50\%$ volume fraction of solids 
(i.e., a remnant volume fraction of $R_r^3 = 0.5$) 
will completely evaporate (at $t^* = 1$)
in $1.24$ milliseconds, according to (\ref{dimensional-time})
with parameters in Table 1. 
But for a butanol droplet of the same size with the same volume fraction
of solids, the time for complete evaporation will be
$1.43$ milliseconds.
Increasing the droplet size by a factor of 2 will increase the 
dimensional time 
by a factor of 4
for droplets of the same composition, 
as indicated by (\ref{dimensional-time}).
If the remnant volume fraction $R_r^3$ is reduced to $0.2$
(or $0.1$),
the time for complete evaporation of pure 
solvent-solid droplets would increase by
a factor of $1.78$ (or $2.12$) from that for $R_r^3 = 0.5$.

Strictly speaking, removal of all the volatile solvent from 
a multicomponent droplet containing nonvolatile cosolvent 
would take forever, i.e., $t^* \to \infty$.  
However, it is still of practical interest to evaluate the time for 
``almost'' complete evaporation. 
Therefore, a time denoted as $t^*_{101}$
for $R(t)$ to reach $1.01 \times R_e$ 
is used here for evaluating the (finite) time $t^*$ needed 
for reaching $R \approx R_e$.  
The value of $t^*_{101}$ can be calculated by 
substituting $1.01 \times R_e$ for $R$ in (\ref{dropRvst}).
For example, 
the values of $t^*_{101}$ for 
a water-EG droplet 
with nonvolatile cosolvent volume fraction $R_n^3 = 0.4$ 
and remnant volume fraction $R_r^3 = 0.5$
at $p_{v, \infty}/p*_{sv, \infty} = 0$ and $0.2$
are $1.23$ and $1.71$, respectively.
But for a butanol-EG droplet 
with $R_n^3 = 0.4$ and $R_r^3 = 0.5$
at $p_{v, \infty}/p*_{sv, \infty} = 0$ and $0.2$,
the values of $t^*_{101}$ become 
are $4.81$ and $5.49$, respectively.
Thus, it takes a water-EG droplet in Fig. 1
with an initial radius of one micron ($R_0 = 10^{-6}$ m)
about ($t^*_{101} \times 1.24$) 
$1.53$ ms and $2.12$ ms to ``almost'' complete 
the evaporation process
at $p_{v, \infty}/p*_{sv, \infty} = 0$ and $0.2$,
whereas a butanol-EG droplet (of $R_0 = 10^{-6}$ m)
in Fig. 1 needs
about ($t^*_{101} \times 1.43$) 
$6.88$ ms and $7.85$ ms. 
Obviously, the presence of the nonvolatile cosolvent 
(with a volume fraction of $R_n^3 = 0.4$) 
hinders the volatile solvent evaporation of a butanol-EG droplet 
much more significantly than that of a water-EG droplet.
This is expected from the fact that 
the value of $B/A$
for the butanol-EG droplet ($= 0.5217$)
is much greater than that
for the water-EG droplet ($= 0.0358$),
at $p_{v, \infty}/p*_{sv, \infty} = 0$. 
In the case of $p_{v, \infty}/p*_{sv, \infty} = 0.2$,
the values of the equilibrium radius $R_e$ become 
$0.810$ for the water-EG droplet and
$0.872$ for the butanol-EG droplet, respectively. 

\begin{figure}[t!] \label{dropEvap2}
\includegraphics[scale=0.7]{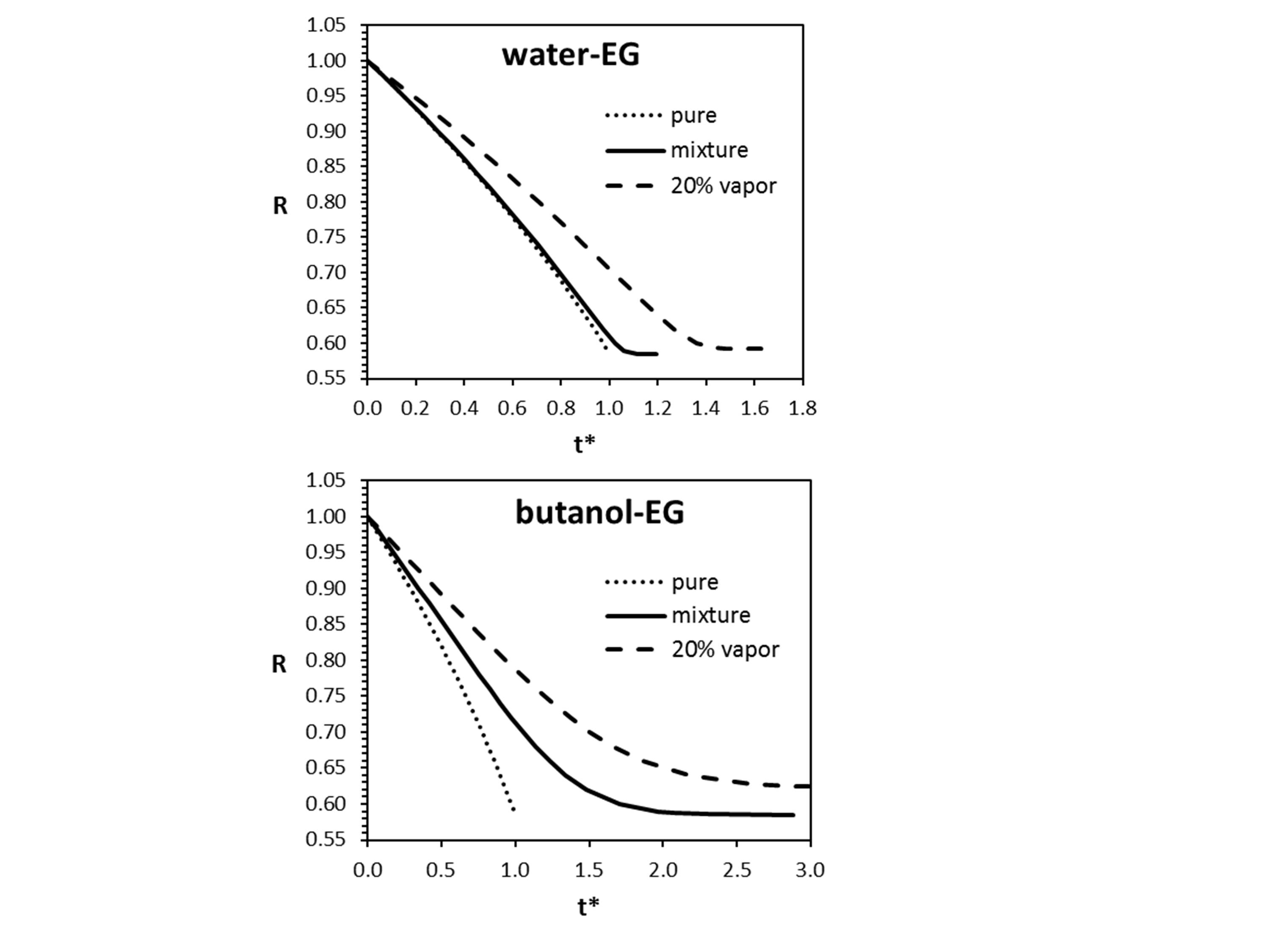}
\caption{As in Fig. 1 but for 
$R_n^3 = 0.1$ and $R_r^3 = 0.2$,
at $p_{v, \infty}/p*_{sv, \infty} = 0$ and $0.2$,
with $t^*_{101} = 1.06$ and $1.37$ for water-EG droplet
while $t^*_{101} = 1.94$ and $2.57$ for butanol-EG droplet.}
\end{figure}

If the volume fraction of nonvolatile solvent 
is reduced to $R_n^3 = 0.1$
with the remnant volume fraction $R_r^3 = 0.2$,
Fig. 2 shows that $R(t)$ for the water-EG droplet
almost coincides with that for $R_n = 0$  
at $p_{v, \infty}/p*_{sv, \infty} = 0$ in a grant scale,
until $R(t)$ becomes very close to $R_e$ ($=0.585$).
Although having $p_{v, \infty}/p*_{sv, \infty} = 0.2$
does not cause much change in the value of $R_e$
(becoming $0.593$ with $x(0) = 0.96$),
it noticeably reduces the evaporation rate for the 
water-EG droplet. 
Despite the substantial reduction of $R_n$ as well as $R_r$,
the butanol-EG droplet still exhibits 
significant nonvolatile cosolvent effect.
Less nonvolatile solvent fraction leads to larger 
$x(0)$ ($= 0.83$).  
Thus, 
$p_{v, \infty}/p*_{sv, \infty} = 0.2$ corresponds to 
$< 25\%$ of vapor equilibrium, which is comparable to 
the case of water-EG droplet shown in Fig. 1.
However, Fig. 2 indicates more significant nonvolatile 
cosolvent effect than that for the water-EG droplet in Fig. 1,
because the value of $B/A$ in this case is $0.130$ (still a few
times larger than $0.0358$ 
for water-EG in Fig. 1), 
at $p_{v, \infty}/p*_{sv, \infty} = 0$. 
For the case of $p_{v, \infty}/p*_{sv, \infty} = 0.2$,
$R_e$ of the butanol-EG droplet
becomes $0.622$ as the final equilibrium droplet radius. 
With $R_n^3 = 0.1$ and $R_r^3 = 0.2$
at $p_{v, \infty}/p*_{sv, \infty} = 0$ and $0.2$,
the values of $t^*_{101}$ for 
a water-EG droplet 
are $1.06$ and $1.37$, respectively.
But for a butanol-EG droplet 
they become 
are $1.94$ and $2.57$, respectively.
Thus, it takes a water-EG droplet in Fig. 2
with an initial radius of one micron ($R_0 = 10^{-6}$ m)
about ($t^*_{101} \times 1.24 \times 1.78$) 
$2.34$ ms and $3.02$ ms to ``almost'' complete 
the evaporation process
at $p_{v, \infty}/p*_{sv, \infty} = 0$ and $0.2$,
whereas a butanol-EG droplet (of $R_0 = 10^{-6}$ m) in Fig. 2 needs
about ($t^*_{101} \times 1.43 \times 1.78$) 
$4.94$ ms and $6.54$ ms.
Interestingly, the reduction of nonvolatile cosolvent 
volume fraction causes a decrease of 
time needed for almost complete evaporation of 
the butanol-EG droplet (comparing that in Fig. 1), 
while the water-EG droplet needs more time,
due to intricate effects of 
the factor of $1.78$ as a consequence of $(1 - R_r^2)$ 
with a reduction of the value of $t^*_{101}$ 
for smaller nonvalotile cosolvent volume fraction $R_n^3$. 

As an apparent general trend shown in Fig.1 and more 
noticeably so in Fig. 2,
the radius of evaporating multicomponent droplet tends to follow the 
``$d^2$ law'' with 
increasing magnitude of slope initially when $R \sim 1$, 
and then has the magnitude of slope decrease gradually 
with $d R/dt \to 0$ as $R \to R_e$.
In other words, evaporating droplet size change is accelerated in
the beginning and then decelerated toward the end,
if a nonvolatile cosolvent is present.
Such an acceleration-deceleration reversal behavior 
is unique in the presence of nonvolatile cosolvent in the 
multicomponent droplet,
as also noted by previous authors with both experiments and 
numerical computations
\citep[e.g.,][]{widmann1997, wilms2005}.
In contrast,
the $d^2$-law behavior of accelerating size reduction 
all the way to $R = R_e$
is clearly
illustrated by the dotted curves for pure solvent cases,
which is more obvious in
Fig. 2 for the case of $R_n^3 = 0.1$.

To have a closer examination of the conditions 
for acceleration-deceleration reversal phenomenon,
it is straightforward to derive 
the formula for $d^2R/dt^2$ from (\ref{dRdt3}) as
\begin{equation} \label{d2Rdt2}
\frac{d^2R}{dt^2} = - \frac{
(1 - {p_{v, \infty}}/p*_{sv, \infty})^2 (R^3 - R_e^3)}
{\pi^2 R^2 \left[A \left(R^3 - R_e^3\right) + B\right]^3}
\left[A\left(R^3 - R_e^3\right)^2 
-2B\left(R^3 - R_e^3\right) - 3 B R_e^3\right] 
\, \, ,
\end{equation}
which indicates an inflexion point ($d^2R/dt^2 = 0$) 
for $R^3 > R_e^3$ at $R = R_i$ with
\begin{equation} \label{Ri}
R_i^3 = R_e^3 + \frac{B}{A}
\left(\sqrt{1+\frac{3AR_e^3}{B}} - 1\right)
\, \, .
\end{equation}
Thus, we have $d^2R/dt^2 < 0$ for $R > R_i$ and 
$d^2R/dt^2 > 0$ for $R < R_i$ with $d^2R/dt^2 \to 0$ as $R \to R_e$.
In an explicit form (\ref{Ri}) suggests that
the value of $R_i$ increases with $B$ which increases with
$p_{v, \infty}/p*_{sv, \infty}$, among others.
It is theoretically possible for $R_i$ to become greater than $1$.
When that happens, the droplet evaporation process 
would not exhibit 
the acceleration-deceleration reversal behavior;
the droplet radius reduction process should show a
decelerating rate all the way to the end.
An evaluation of $R_i$ according to (\ref{Ri}) shows that
$R_i = 0.887$ and $0.929$ at 
$p_{v, \infty}/p*_{sv, \infty} = 0$ and $0.2$ for 
the water-EG droplet in Fig. 1, which indeed exhibits
the acceleration-deceleration reversal behavior (because $R_i < 1$).
But for the butanol-EG droplet in Fig. 1,
which do not seems to show obvious accleration-deceleration
reversal,
the values of $R_i$ are $1.002$ and $1.101$ 
(both are greater than $1$) at
$p_{v, \infty}/p*_{sv, \infty} = 0$ and $0.2$. 
With a reduction of $B$ via reduction of the nonvolatile 
cosolvent volume fraction $R_n^3$ to $0.1$ as in Fig. 2
(for $R_r^3 = 0.2$, where the acceleration-deceleration reversal
is observable),
the butanol-EG droplet would have 
$R_i = 0.723$ and $0.773$ at
$p_{v, \infty}/p*_{sv, \infty} = 0$ and $0.2$. 

\subsection{Condensational Growth of Droplet}
Droplet can grow by condensation of the volatile solvent vapor 
when $p_{v, \infty}/p*_{sv, \infty} > x(0)$, which corresponds to  
$R_e > 1$ if 
$p_{v, \infty}/p*_{sv, \infty} \le 1$, 
again until $R$ reaches $R_e$.
According to (\ref{Rr3}), 
$R_e \to \infty$ at 
the limit of 
$p_{v, \infty}/p*_{sv, \infty} = 1$, 
suggesting that the droplet can grow indefinitely without bound. 
However, for
$p_{v, \infty}/p*_{sv, \infty} > 1$,
the negative value of 
$R_e$ loses physical meaning, 
as it monotonically increases 
from $- \infty$ 
with $p_{v, \infty}$ increasing from 
$p*_{sv, \infty}$.
It should be noted that (\ref{Rr3})
can be rewritten as 
$p_{v, \infty}/p*_{sv, \infty}$ 
$= \{1 + \rho_n M_v R_n^3/[\rho_v M_n (R_e^3 - R_r^3)]\}^{-1}$, 
which can only relate to the volatile solvent mole fraction $x$ when
$p_{v, \infty}/p*_{sv, \infty} \le 1$, 
corresponding to $R_e \ge R_r$.
In that case, 
the droplet size will change until $R = R_e$ 
when the volatile solvent mole fraction $x(t)$ in the droplet
reaches the thermodynamic equilibrium value of 
$p_{v, \infty} / p*_{sv, \infty}$.

\begin{figure}[t!] \label{dropEvap3}
\includegraphics[scale=0.7]{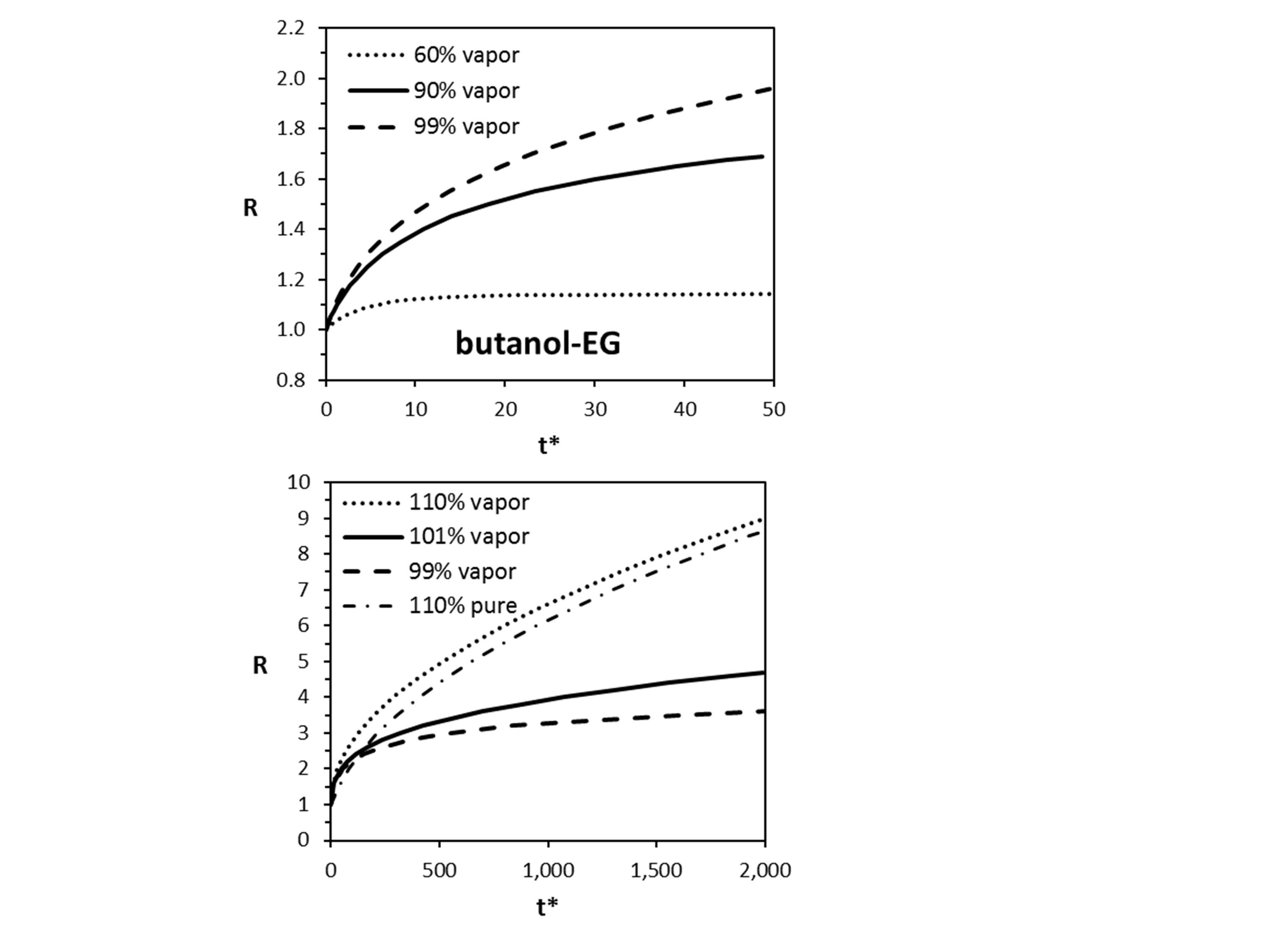}
\caption{As in Fig. 1 
(with nonvolatile cosolvent volume fraction $R_n^3 = 0.4$ 
and remnant volume fraction $R_r^3 = 0.5$)
but only for butanol-EG droplet
at $p_{v, \infty}/p*_{sv, \infty} = 0.6$, $0.9$, $0.99$, $1.01$, and $1.10$
to illustrate characteristics of condensation growth of 
a multicomponent droplet. 
The dot-dash curve labeled ``110\% pure'' is 
for a reference case 
with $R_n^3 = 0$ and $R_r^3 = 0.5$ for pure butanol droplet
at $p_{v, \infty}/p*_{sv, \infty} = 1.10$.
The values of $t^*_{099}$ at 
$p_{v, \infty}/p*_{sv, \infty} = 0.6$, $0.9$, and $0.99$ are
$13$, $172$, and $8131$. 
The values of $t^*_5$ at 
$p_{v, \infty}/p*_{sv, \infty} = 1.01$ and $1.10$ are
$2488$, and $514$.
For the dot-dash curve with $R_n = 0$, the value of 
$t^*_5$ is $649$.} 
\end{figure}
 
The condensation growth behavior of a butanol-EG droplet is 
illustrated in Fig. 3, 
with nonvolatile cosolvent volume fraction $R_n^3 = 0.4$ and 
remnant volume fraction $R_r^3 = 0.5$,
at $p_{v, \infty}/p*_{sv, \infty} = 0.6$, $0.9$, $0.99$, $1.01$, and $1.10$.
Because the value of $x(0)$ is $0.432$
for $R_n^3 = 0.4$ and $R_r^3 = 0.5$,
condensation growth will occur whenever 
$p_{v, \infty}/p*_{sv, \infty} > 0.432$.
The values of $R_e$ corresponding to  
$p_{v, \infty}/p*_{sv, \infty} = 0.6$, $0.9$, and $0.99$ are
$1.14$, $1.86$, and $4.03$, respectively.  
Thus, when
$p_{v, \infty}/p*_{sv, \infty} < 1$,
condensation growth of a multicomponent droplet 
will practically reach a finite size $R_e$ (as $t^* \to \infty$).
In other words, a multicomponent droplet cannot grow indefinitely
without bound in an environment of 
$p_{v, \infty}/p*_{sv, \infty} < 1$ by vapor condensation.
Useful insights may be gained, however, by examining
the normalized time $t^*_{099}$ for $R$ to reach $0.99 \times R_r$,
similar to $t^*_{101}$ for evaporating droplet. 
For example,
with $R_n^3 = 0.4$ and $R_r^3 = 0.5$,
the values of $t^*_{099}$
are $13$, $172$, and $8131$, for 
$p_{v, \infty}/p*_{sv, \infty} = 0.6$, $0.9$, and $0.99$,
respectively.
Thus, a butanol-EG droplet of $R_0 = 10^{-6}$ m will take 
$0.029$, $0.385$, and $18.295$ seconds to reach $0.99 \times R_r$
for
$p_{v, \infty}/p*_{sv, \infty} = 0.6$, $0.9$, and $0.99$, 
given that $t^* = 1$ corresponding to $2.25$ milliseconds.
When 
$p_{v, \infty}/p*_{sv, \infty} \ge 1$, 
a multicomponent droplet will grow indefinitely with time,
as seen in Fig. 3 for 
$p_{v, \infty}/p*_{sv, \infty} = 1.01$ and $1.10$.
In this case, the normalized time $t^*_5$ for $R = 5$ 
might be a useful measure as the time needed for a droplet
to grow $125$ times of its initial volume. 
For a butanol-EG droplet
with $R_n^3 = 0.4$ and $R_r^3 = 0.5$,
the values of $t^*_{5}$
are $2488$ and $514$, for 
$p_{v, \infty}/p*_{sv, \infty} = 1.01$ and $1.10$,
respectively.
As a reference, a dot-dash curve is also shown in Fig. 3
for a pure butanol droplet
(with $R_n^3 = 0$ and $R_r^3 = 0.5$)   
at $p_{v, \infty}/p*_{sv, \infty} = 1.10$,
indicating slower growth of $R(t)$ than a butanol-EG droplet
especially near $t^* = 0$.
The value of $t^*_5$ for the dot-dash curve with $R_n = 0$ 
in Fig. 3 is $649$.
But with time, the dotted curve will converge to 
the dot-dash curve, because as the droplet grows indefinitely its 
volatile solvent mole fraction will increase toward unity 
asymptotically.   

\begin{figure}[t!] \label{dropEvap4}
\includegraphics[scale=0.75]{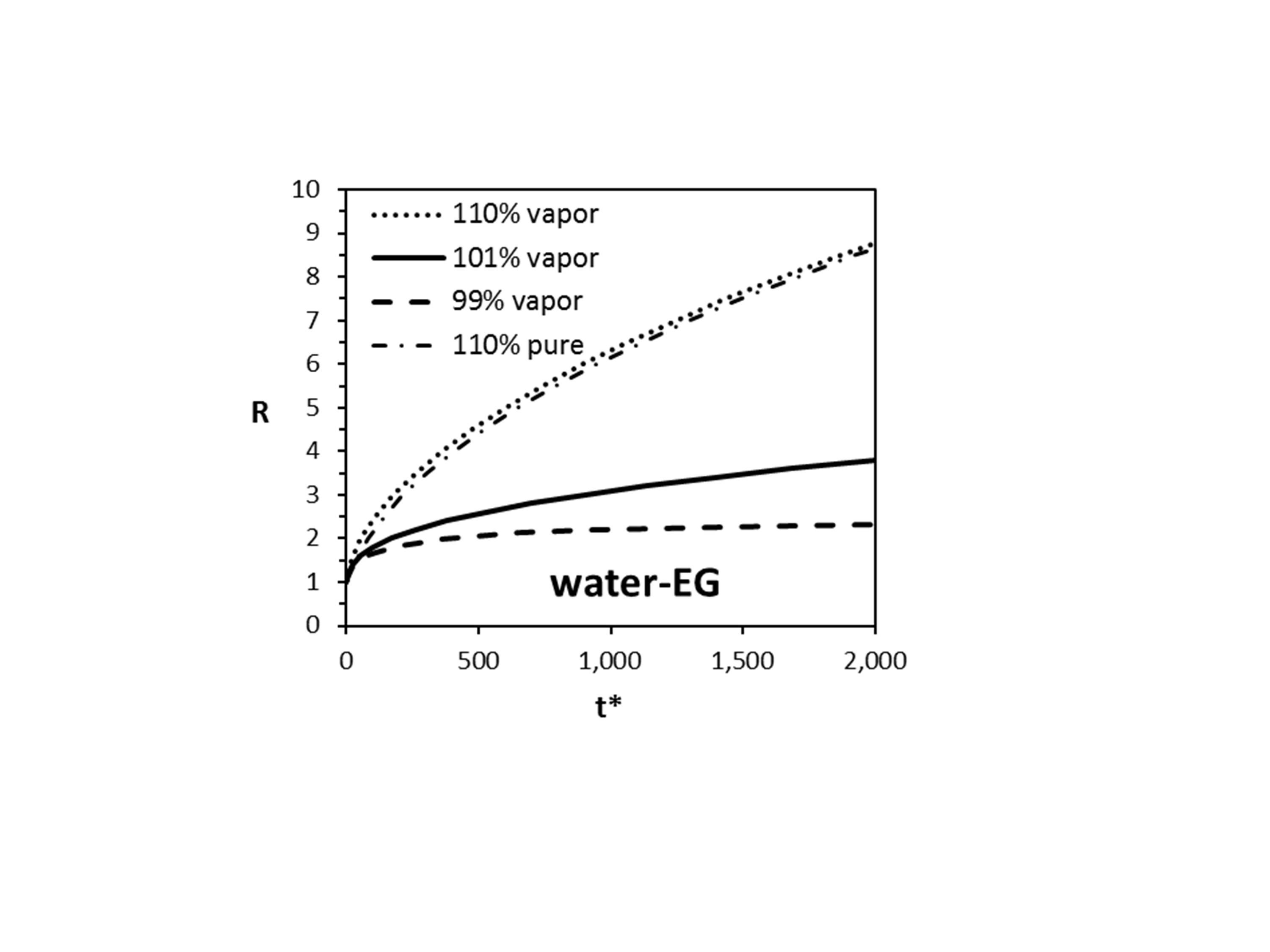}
\caption{As in Fig. 3 but for water-EG droplet
at $p_{v, \infty}/p*_{sv, \infty} = 0.99$, $1.01$, and $1.10$.
The value of $t^*_{099}$ at 
$p_{v, \infty}/p*_{sv, \infty} = 0.99$ is
$2660$. 
The values of $t^*_5$ at 
$p_{v, \infty}/p*_{sv, \infty} = 1.01$ and $1.10$ are
$4471$, and $600$.
For the dot-dash reference curve with $R_n = 0$, the value of 
$t^*_5$ is $649$.} 
\end{figure}

Similar to Fig. 3, 
the condensation growth behavior of water-EG droplet
is shown in Fig. 4. 
Because it is necessary to have
$p_{v, \infty}/p*_{sv, \infty} > 0.795$
($= x(0)$, which is larger than that for the butanol-EG mixture) 
for condensation growth to occur,
a water-EG droplet is expected to grow at a slower rate than a 
butanol-EG droplet as reflected in a comparison of Fig.3 and Fig. 4.
For the water-EG droplet
with $R_n^3 = 0.4$ and $R_r^3 = 0.5$,
the value of $t^*_{099}$
is $2660$ for 
$p_{v, \infty}/p*_{sv, \infty} = 0.99$,
and the values of $t^*_{5}$
are $4471$ and $600$, for 
$p_{v, \infty}/p*_{sv, \infty} = 1.01$ and $1.10$,
respectively.
The value of $t^*_5$ for the dot-dash curve with $R_n = 0$ 
in Fig. 4 is $649$ again (as independent of solvent type
in the absence of nonvolatile cosolvent due to 
the definition of normalized time).
A less significant difference 
between curves for the water-EG droplet (dotted) and 
pure water droplet (dot-dash) also suggests  
a weaker nonvolatile cosolvent effect on vapor transport 
for the water-EG droplet
than for butanol-EG droplet, as noticed in 
the case of evaporating droplet.

In general, condensation growth of droplet exhibits 
shallower slope than evaporating droplet, 
as illustrated by comparing time scales shown in Fig. 1, Fig. 2 with 
that in Fig. 3, Fig.4.
This may be explained by the fact that 
$d R/dt \sim 1/R$ according to (\ref{dRdt3}), 
indicating a tendency of accelerating droplet size reduction 
during evaporation while decelerating droplet size growth
during condensation. 
In the case of condensation growth, the volatile solvent 
mole fraction increases with time and therefore
the nonvolatile cosolvent effect decreases progressively in the process.
Thus, no physical mechanism exists for 
the acceleration-deceleration reversal phenomenon 
in condensational growth of a multicomponent droplet.
While the presence of nonvolatile cosolvents tends to hinder
droplet evaporation,
it consistently enhances the condensational growth rate of 
a multicomponent droplet due to the effective reduction of 
saturation vapor pressure of the volatile solvent.

\begin{figure}[t!] \label{TR}
\includegraphics[scale=0.75]{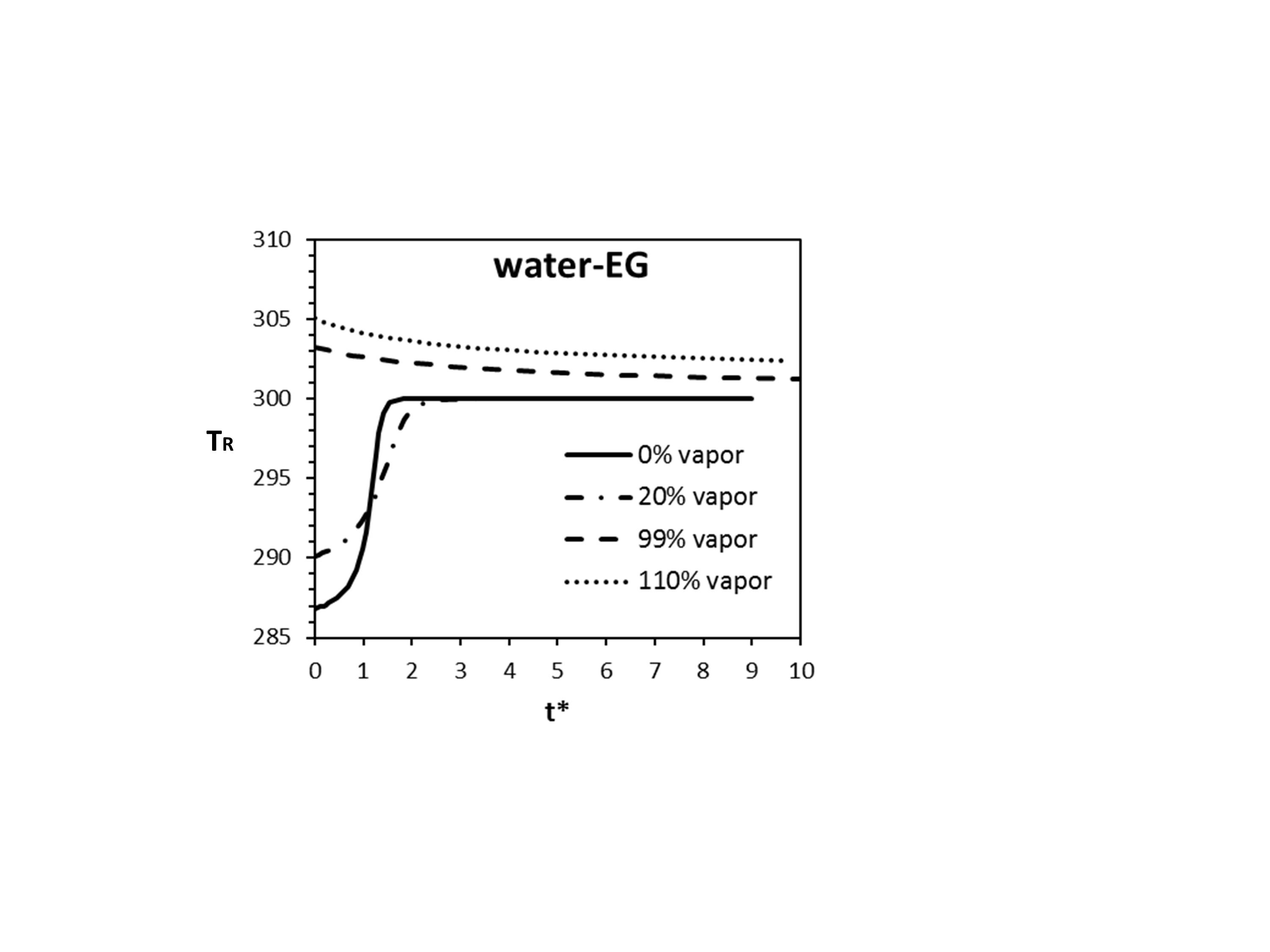}
\caption{Surface temperature $T_R$ evolution for water-EG droplet
with nonvolatile cosolvent volume fraction $R_n^3 = 0.4$ 
and remnant volume fraction $R_r^3 = 0.5$,
at $p_{v, \infty}/p*_{sv, \infty} = 0$ (solid), 
$0.2$ (dot-dash), $0.99$ (dashed), and $1.10$ (dotted)
according to (\ref{T_R}).}
\end{figure}

\subsection{Droplet Surface Temperature}
In theoretical analysis, 
an explicit expression of the local temperature at droplet surface
is sometimes useful \citep[e.g.,][]{kulmala1993, vesala1997}.
As a by-product of the mathematical derivation,
the formula for
local temperature at the multicomponent droplet surface $T_R$, 
according to (\ref{qssEnergy}) and (\ref{rhodiff}),
can be obtained as
\begin{equation} \label{T_R}
T_R = T_{\infty}\left(1 + \frac{\pi \rho_l L D}{2 k T_{\infty}} 
\frac{d R^2}{dt}\right)
= T_{\infty} \left[1 - \frac{\rho_l L D}{k T_{\infty}}
\frac{1 - p_{v, \infty}/p*_{sv, \infty}}{A + B/(R^3 - R_e^3)}\right]
\, \, ,
\end{equation}
with $d R^2/dt$, $R_e^3$, $A$, and $B$ given by (\ref{dRdt3})---(\ref{Bdef}).
Because $B = 0$ when $R_n = 0$, droplet surface temperature $T_R$ 
remains constant because 
$d R^2/dt = constant$  
in the absence of the nonvolatile cosolvent.

For a water droplet ($R_n = 0$)
at $p_{v, \infty}/p*_{sv, \infty} = 0$, $0.2$, $0.99$, and $1.10$
with $T_{\infty} = 300$ K,
its surface temperature $T_R = 285.87$, $288.69$, $299.86$, and $301.41$ K,
respectively.
The presence of nonvolatile cosolvent leads to 
at time-dependent $T_R$ for $d R^2/dt$ is not a constant anymore. 
Fig. 5 shows the curves of $T_R(t^*)$ for water-EG droplet 
with $R_n^3 = 0.4$ and $R_r^3 = 0.5$
at $p_{v, \infty}/p*_{sv, \infty} = 0$ (solid), $0.2$ (dot-dash),
$0.99$ (dashed), and $1.10$ (dotted).
For $p_{v, \infty}/p*_{sv, \infty} < 1$,
each curve approaches the asymptotic value $300$ K as
$d R^2/dt \to 0$ with increasing $t^*$, 
because the diminishing rate for vapor transport leads to
an isothermal process.  
For $p_{v, \infty}/p*_{sv, \infty} = 1.10$ ($> 1$)
with $R_n^3 = 0.4$ and $R_r^3 = 0.5$
(dotted curve in Fig. 5),
the value of $T_R$ asymptotically approaches $301.41$ K ($> 300$ K) 
due to a non-diminishing 
$d R^2/dt$ 
even as $R \to \infty$ where 
$d R^2/dt \to 2 (
p_{v, \infty}/p*_{sv, \infty} - 1)/(\pi A)$.
This phemomanon is generally true for
$p_{v, \infty}/p*_{sv, \infty} > 1$
according to (\ref{dRdt3}).
The variation of surface temperature $T_R$ 
appears to be much more dramatic for evaporating droplet 
than growing droplet by condensation. 
The ``S-shape'' curves in Fig. 5
for evaporating multicomponent droplet 
are consistent with the acceleration-deceleration reversal
behavior exhibited in $R(t^*)$ curves in Fig. 1 and Fig. 2.
Without the acceleration-deceleration reversal,
the surface temperature evolution curves $T_R(t^*)$ for 
a droplet during condensational growth appear less eventful.

\section{Summary}
Analytical formulas are derived in the present work for 
evaluating vapor transport of a volatile solvent 
for an isolated multicomponent droplet 
in a quiescent environment,
based on quasi-steady-state approximation.
Among multiple solvent components, only one component is considered to be
much more volatile than the rest such that other components are 
assumed to remain unchanged in the droplet during the process of 
volatile solvent evaporation or condensation.
With direct application to the
Aerosol Jet$^{\circledR}$
printing in mind, simplifications were made justifiably for
ink droplet diameter around a range of $1$ to $5$ $\mu$m 
such as ignoring the Knudsen number effect, 
negligible vapor mass concentration compared with the ambient gas density,
as well as applicability of Raoult's law for an ideal solution 
of multicomponent solvent mixture.
The form of derived equation for $d R^2/dt$ 
suggests that in the presence of nonvolatile cosolvent,
the rate of droplet size change diminishes
as the amount of volatile solvent in the droplet 
approaches equilibrium value with the ambient vapor pressure
($R \to R_e$).
In the absence of nonvolatile cosolvent, 
the droplet radius $R$ changes according to the 
familier ``$d^2$ law'', which is recovered from
the general formula derived herewith when 
the volume fraction of nonvolatile cosolvent 
becomes zero ($R_n = 0$). 		

In the case of evaporating droplet, 
the radius $R(t^*)$ initially follows the 
``$d^2$ law'' near $R(0) = 1$ with an accelerated rate of change. 
However, the presence of 
nonvolatile cosolvent is predicted to continuously reduce the rate of 
evaporation as the mole fraction of the volatile solvent 
keeps decreasing.  According to Raoult's law, decreasing the 
mole fraction effectively corresponds to a reduction of 
saturation vapor pressure for the volatile solvent and thereby 
weakens the driving force for evaporation. 
Thus, the magnitude of droplet size change is eventually 
decelerated toward the end with $d R/dt \to 0$ as $R \to R_e$.
Such an acceleration-deceleration reversal behavior in
the droplet size change is unique in the presence of 
nonvolatile cosolvent, 
while the droplet of pure solvent follows the 
$d^2$ law of accelerating size reduction 
all the way through the end. 
A closer examination of the explicit formula for $d^2R/dt^2$,
however, 
indicates that the acceleration phase may disappear 
when the nonvolatile cosolvent volume fraction $R_n^3$ is small
and the ambient vapor pressure $p_{v, \infty}$ is 
relatively high.

Because the net effect of adding nonvolatile cosolvent is 
to reduce the mole fraction of the volatile solvent such that
the saturation vapor pressure is lowered, 
vapor condensation onto the multicomponent droplet is predicted to 
occur when the ambient vapor pressure is subsaturated 
with respect to that for the pure volatile solvent. 
In this case, the droplet will grow asymptotically toward a 
finite size $R = R_e$ ($> 1$).
Such restricted growth of droplet by condensation is 
unique only when the nonvolatile cosolvent is present 
in the multicomponent droplet.
But when the ambient vapor pressure becomes supersaturated 
with respect to that for the pure volatile solvent,
the condensation growth of droplet can continue indefinitely 
without bound.  Condensation growth of droplet 
generally exhibits shallower slope than evaporating droplet 
because the droplet size change is 
continuously decelerated with time as illustrated in
figures of $R(t^*)$ and $T_R(t^*)$.

The effects of nonvolatile cosolvent on 
the volatile solvent evaporation-condensation characteristics
predicted here should occur with 
general multicomponent droplets,
in view of the fact that most multicomponent mixtures contain
cosolvents with different volatilities.
It is expected more often than not to observe 
a single component of the most volatile solvent 
evaporating or condensing 
at a much shorter time scale than the other cosolvents.
Thus, the formulas derived here may become applicable to 
more general situations with multicomponent droplets
as long as the evaporation-condensation time scales 
for different components differ substantially.

\section*{Acknowledgments}
The author would like to thank John Lees for 
encouragement and guidance, and Dr. Mike Renn,
Dr. Kurt Christenson, Jason Paulsen, as well as 
many other Optomec colleagues, for 
helpful technical discussions.

\section*{Figure Captions}

\vspace{4 mm}

{Figure 1:  Evaporating droplet radius evolution with time
for water-EG and butanol-EG mixtures,
with ethylene glycol (EG) serving as the nonvolatile
cosolvent with a volume fraction of
$R_n^3 = 0.4$.
The remnant volume of droplet (excluding that of 
the volatile solvent) is denoted as $4 \pi R_r^3/3$
here in this figure with the remnant volume fraction $R_r^3 = 0.5$.
The solid curve and dashed curve correspond to 
$p_{v, \infty}/p*_{sv, \infty} = 0$ and $0.2$,
with $t^*_{101} = 1.23$ and $1.71$ for water-EG droplet
while $t^*_{101} = 4.81$ and $5.49$ for butanol-EG droplet.
The dotted curve is for pure volatile solvent with 
$R_n^3 = 0$ and $R_r^3 = 0.5$ at
$p_{v, \infty}/p*_{sv, \infty} = 0$, as a reference.}

\vspace{4 mm}

\noindent
{Figure 2:  As in Fig. 1 but for 
$R_n^3 = 0.1$ and $R_r^3 = 0.2$,
at $p_{v, \infty}/p*_{sv, \infty} = 0$ and $0.2$,
with $t^*_{101} = 1.06$ and $1.37$ for water-EG droplet
while $t^*_{101} = 1.94$ and $2.57$ for butanol-EG droplet.}

\vspace{4 mm}

\noindent
{Figure 3:  As in Fig. 1 
(with nonvolatile cosolvent volume fraction $R_n^3 = 0.4$ 
and remnant volume fraction $R_r^3 = 0.5$)
but only for butanol-EG droplet
at $p_{v, \infty}/p*_{sv, \infty} = 0.6$, $0.9$, $0.99$, $1.01$, and $1.10$
to illustrate characteristics of condensation growth of 
a multicomponent droplet. 
The dot-dash curve labeled ``110\% pure'' is 
for a reference case 
with $R_n^3 = 0$ and $R_r^3 = 0.5$ for pure butanol droplet
at $p_{v, \infty}/p*_{sv, \infty} = 1.10$.
The values of $t^*_{099}$ at 
$p_{v, \infty}/p*_{sv, \infty} = 0.6$, $0.9$, and $0.99$ are
$13$, $172$, and $8131$. 
The values of $t^*_5$ at 
$p_{v, \infty}/p*_{sv, \infty} = 1.01$ and $1.10$ are
$2488$, and $514$.
For the dot-dash curve with $R_n = 0$, the value of 
$t^*_5$ is $649$.} 

\vspace{4 mm}

\noindent
{Figure 4:  As in Fig. 3 but for water-EG droplet
at $p_{v, \infty}/p*_{sv, \infty} = 0.99$, $1.01$, and $1.10$.
The value of $t^*_{099}$ at 
$p_{v, \infty}/p*_{sv, \infty} = 0.99$ is
$2660$. 
The values of $t^*_5$ at 
$p_{v, \infty}/p*_{sv, \infty} = 1.01$ and $1.10$ are
$4471$, and $600$.
For the dot-dash reference curve with $R_n = 0$, the value of 
$t^*_5$ is $649$.} 

\vspace{4 mm}

\noindent
{Figure 5:  Surface temperature $T_R$ evolution for water-EG droplet
with nonvolatile cosolvent volume fraction $R_n^3 = 0.4$ 
and remnant volume fraction $R_r^3 = 0.5$,
at $p_{v, \infty}/p*_{sv, \infty} = 0$ (solid), 
$0.2$ (dot-dash), $0.99$ (dashed), and $1.10$ (dotted)
according to (\ref{T_R}).}


\end{document}